\documentclass[prb,twocolumn,amsmath,showpacs]{revtex4}

\usepackage{epsfig}
\usepackage{graphicx}

\begin{document}
\input epsf

\title {Detailed magnetization study of superconducting properties of YBa$_2$Cu$_3$O$_{7-x}$ ceramic spheres}

\author {I. L. Landau, J. B. Willems and J. Hulliger}
\affiliation{Universit\"at Bern, Department f\"ur Chemie und Biochemie, Freiestrasse 3, CH-3012 Berne, Switzerland}

\date{\today}

\begin{abstract}

We present a magnetization study of low density YBa$_2$Cu$_3$O$_{7-x}$ ceramics carried out in magnetic fields 0.5 Oe $<H<$ 50 kOe. It was demonstrated that superconducting links between grains may be completely suppressed either by a magnetic field $H \sim 100$ Oe (at low temperatures) or by an increase of temperature above 70 K. This property of present samples allowed to evaluate the ratio between an average grain size and the magnetic field penetration depth $\lambda$. Furthermore, at temperatures $T > 85$ K, using low-field magnetization measurements, we could evaluate the temperature dependence of $\lambda$, which turned out to be very close to predictions of the conventional Ginzburg-Landau theory. Although present samples consisted of randomly oriented grains, specifics of magnetization measurements allowed for evaluation of $\lambda_{ab}(T)$. Good agreement between our estimation of the grain size with the  real sample structure provides evidence for the validity of this analysis of magnetization data. Measurements of equilibrium magnetization in high magnetic fields were used for evaluation of $H_{c2}(T)$. At temperatures close to $T_c$, the $H_{c2}(T)$ dependence turned out to be linear in agreement with the Ginzburg-Landau theory. The value of temperature, at which $H_{c2}$ vanishes, coincides with the superconducting critical temperature evaluated from low-field measurements.

\end{abstract}

\maketitle

\section{Introduction}

In this work we present a detailed magnetization study of low density ceramics of YBa$_2$Cu$_3$O$_{7-x}$ (Y-123). Data were collected in magnetic fields $0.5$ Oe $\le H \le 50$ kOe. The main attention was paid to the analysis of reversible magnetization, which provides information on equilibrium properties of the superconducting state. 

It turned out that superconducting links between the grains are weak and can be suppressed either by a magnetic field as low as 100 Oe (at low temperatures) or by increasing temperature above $T = 70$ K. As we discuss below, this provides the possibility to study a system of non-interacting grains for gaining the information about the relation between the grain size and the magnetic field penetration depth $\lambda$. 

In magnetic fields $H \sim 1$ Oe, an average grain size of 3-5 $\mu$m) is too small to capture even a single vortex line, which makes the low-field temperature dependence of the sample magnetization $M(T)$ reversible. Analyzing the reversible $M(T)$ curve, the temperature dependence of the magnetic field  penetration depth $\lambda$ can be obtained. Although present samples consisted of randomly oriented grains, specifics of magnetization measurements allowed for an evaluation of $\lambda_{ab}(T)$. Quite surprisingly, $M(T)$ data at $T \gtrsim 85$ K follow very closely predictions of the conventional Ginzburg-Landau theory.

Magnetization measurements in fields $H \gtrsim 1$ kOe probe completely different physics. In this case, the mixed state is established inside grains and magnetization data provide some of its characteristics. Also here we were interested in reversible magnetization data, which reflect equilibrium properties of the mixed state. As was demonstrated in Ref. \onlinecite{land1}, equilibrium magnetization $M(H,T)$ data allow to establish the temperature dependence of the normalized upper critical field $H_{c2}$. In this work, we extended measurements up to $T \approx 0.995T_c$ and it was demonstrated that at temperatures close to $T_c$ the $H_{c2}(T)$ dependence is indeed linear in agreement with the Ginzburg-Landau theory. Linearity of $H_{c2}(T)$ also allows for a rather accurate evaluation of $T_c$. We emphasize that $T_c$ determined in such a way is in perfect agreement with the results of low field $M(T)$ measurements. We consider this agreement as additional evidence that the scaling procedure developed in Ref. \onlinecite{land1} may serve as a reliable tool for the analysis of magnetization data.

\section{Experimental}

The samples were made from commercial YBa$_2$Cu$_3$O$_{7-x}$ powder (Alfa Aesar). About 0.6 g of powder was suspended in 6 g of 2-butanone by vigorous stirring. Small droplets of this suspension were dropped with a syringe into hemispherical (5 mm diameter) templates made in a flat ZrO$_2$/BN plate (IEPCO AG). The substrate was preliminary heated to 250 $^o$C. Rapid evaporation of the solvent lead to flotation of droplets above the surface of the substrate. After several seconds of such a flotation, solid spherical samples were formed.  These spheres were sintered in a preheated furnace at 700 $^o$C for 30 min. Thereafter,  spherical samples were solid enough and could be removed from the substrate and sorted by size. About 30 well formed spheres in the size range of 0.4 to 1.2 mm were finally sintered in a Al$_2$O$_3$ crucible at 940 ¡C (heating rate 300$^o$C/h) in an oxygen atmosphere for 24 h followed by cooling down to room temperature in a rate of 15$^o$C/h.

Two samples from this series (Y\#1, Y\#2) were chosen for measurements. Their shape was very close to an oblate sphere as it is schematically illustrated in the inset of Fig. 1. Both samples had practically identical size $2a = (1.2 \pm 0.02)$ mm and  $2c = (0.86 \pm 0.03)$ mm with $a/b = (1.4 \pm 0.07)$. While the volumes $V = (0.65 \pm 0.04)$ mm$^3$ of the samples were almost the same, the masses were slightly different: $m^{(1)} = (2.4 \pm 0.04)$ mg (Y\#1) and $m^{(2)} = (2.0 \pm 0.04)$ mg (Y\#2). Average densities of the samples were about 0.5 of that for Y-123 single crystals.

All measurements were carried out on a SQUID Magnetometer with a 5 T magnet (Quantum Design).

\section{Experimental results}
 
\begin{figure}[h]
 \begin{center}
  \epsfxsize=1.0\columnwidth \epsfbox {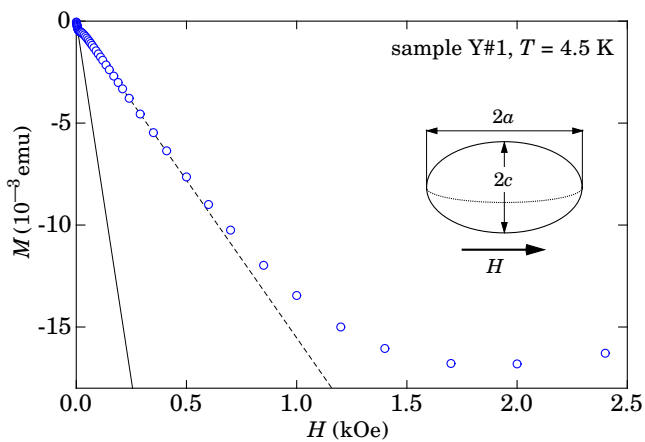}
  \caption{(Color online) $M$ versus $H$ after zero-field cooling to $T = 4.5$ K. A magnetic field was applied parallel to the equatorial plane of the sample as it is illustrated in the inset. The solid and the dashed lines are the best linear fits to the low and the intermediate parts of the $M(H)$ curve (see Figs. 2 and 3).}
 \end{center}
\end{figure}
The low temperature magnetization curves are shown in Figs. 1-3. In very low fields, $M$ is a linear function of $H$ (Fig. 2). The derivative $dM/dH$ is practically the same for both samples. 

\begin{figure}[!h]
 \begin{center}
  \epsfxsize=1.\columnwidth \epsfbox {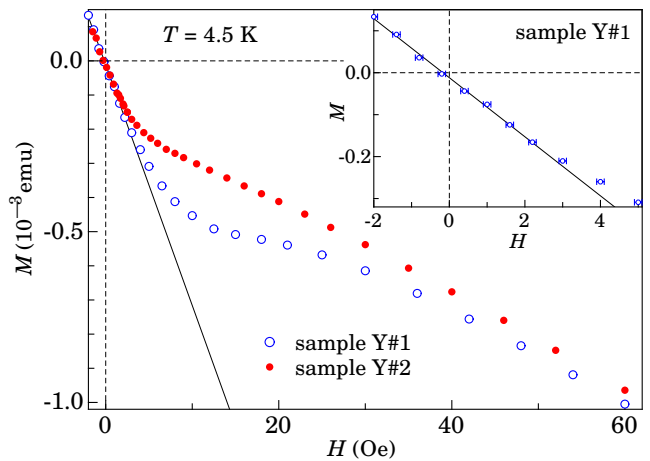}
  \caption{(Color online) Low field parts of the $M(H)$ curves. The inset shows results for the sample Y\#1 on expanded scales. The straight line is the best linear fit to data points  measured in $|H| \le 2$ Oe.}
 \end{center}
\end{figure}

The magnetic moment of a superconducting sample may be written as
\begin{equation}
\frac{1}{V}\frac{4\pi M}{H} = \frac{1}{1 + 4\pi \chi N},
\end{equation}
where $N$ is the demagnetizing factor of a sample and $\chi$ is its magnetic susceptibility. In the ideal Meissner state $4\pi \chi = -1$. Substituting the experimental value of $dM/dH = (7.1 \pm 0.4)\cdot 10^{-5}$ emu/Oe and the sample volume, we obtain $N = 0.27$, which is in very good agreement with $N = 0.29$, calculated for an ellipsoid of the corresponding shape (magnetic field oriented along the equatorial plane).\cite{demag} This agreement shows that in fields of some Oersteds, superconducting links between grains are strong enough to ensure the coherent superconducting state throughout a sample. In slightly higher fields, however, the deviation of the $M(H)$ curves from linearity indicates breaking of superconducting links between grains and penetration of magnetic flux inside the sample. We note that for the less dense sample Y\#2 this penetration starts in even weaker fields reflecting somewhat weaker intergrain coupling (Fig. 2).  

\begin{figure}[h]
 \begin{center}
  \epsfxsize=1.0\columnwidth \epsfbox {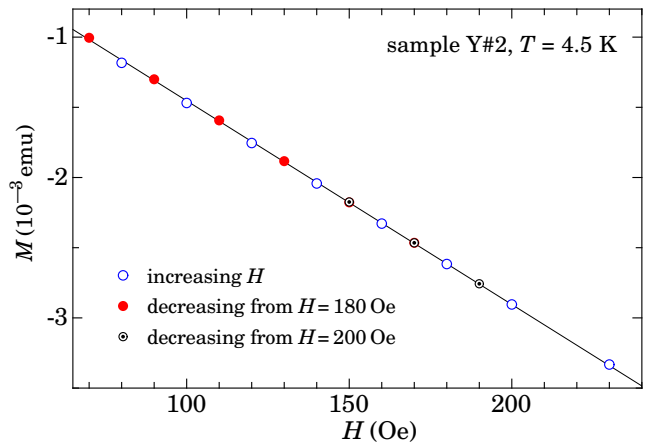}
  \caption{(Color online) An intermediate part of the $M(H)$ curve for the sample Y\#2. The straight line is the best linear fit to data points  collected in magnetic fields between 120 and 200 Oe.}
 \end{center}
\end{figure}
In magnetic fields 50 Oe $< H < 250$ Oe, the $M(H)$ dependence is again close to a linear function (see Figs. 1 and 3). $M(H)$ data for the sample Y\#2 in the corresponding field range are shown in Fig. 3. In addition to its linearity, the $M(H)$ curve is perfectly reversible.  This behavior is a clear sign that superconductivity of intergrain links is completely suppressed by a magnetic field, while the field is not yet strong enough to overcome pinning barriers and to penetrate inside grains. The effective formation of the mixed state in grains starts upon substantially higher magnetic fields $H > 0.5$ kOe, which is indicated by a saturation of the $M(H)$ curve presented in Fig. 1. 

Thus, in magnetic fields between 100 and 200 Oersteds, our samples can be considered as ensembles of non-interacting grains and their magnetic behavior should be similar to that of powder samples. In this magnetic field range, the derivative $dM/dH$ may be compared with a total volume of all superconducting grains, $V_S = m/\rho$, where $m$ is the sample mass and $\rho$ is the bulk density of Y-123. For the sample Y\#2, $V_S$ is 0.31 mm$^3$. Taking an experimental value of $dM/dH = 1.45 \cdot 10^{-5}$ emu/Oe, we get $(1/V_S)(4\pi dM/dH) = 0.58 < 1$, i.e., the volume, from which the magnetic field is expelled, is smaller than $V_S$. This means that the average grain size is comparable with the magnetic field penetration depth $\lambda$. 

\begin{figure}[h]
 \begin{center}
  \epsfxsize=1.0\columnwidth \epsfbox {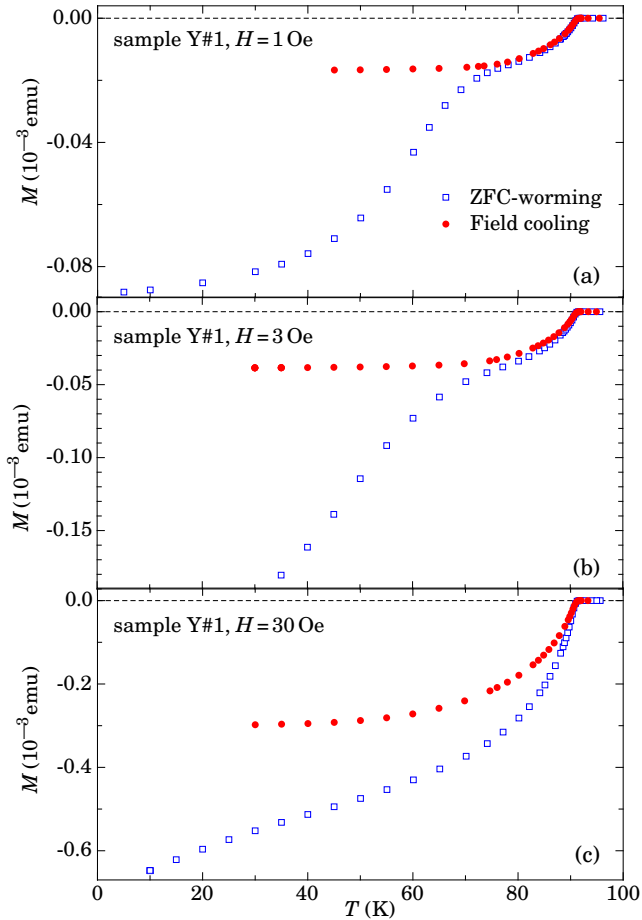}
  \caption{(Color online) Zero-field-cooled (ZFC) and field-cooled (FC) $M(T)$ curves. The applied magnetic field is perpendicular to the equatorial plane of the sample. (a) $H = 1$ Oe. (b) $H = 3$ Oe. (c) $H = 30$ Oe.}
 \end{center}
\end{figure}
Fig. 4 displays temperature dependences of the sample magnetization measured in three different magnetic fields. Considering these plots, we note that in addition to 0.1 Oe uncertainty of a SQUID magnetometer  there is about $-0.2$ Oe field (see the inset of Fig. 2) corresponding to a "zero" applied field. However, we do not use the absolute value of $H$ in our analysis. Subsequently, this uncertainty does not influence the results.   

A steep decline of the sample diamagnetism at $T \approx 60$ K, which may be seen in Fig. 4(a), corresponds to an effective critical temperature of intergrain links. Another interesting feature, which can be seen in Fig. 4(a), is, that at temperatures $T \gtrsim 70$ K and $H = 1$ Oe, the sample magnetization is practically reversible, while this reversibility completely disappears in $H = 30$ Oe (see Fig. 4(c)). In the case of $H = 3$ Oe (Fig. 4(b)), the sample magnetization is close to be reversible although the distance between the two magnetization curves is clearly visible. 

As it is well known, pinning effects are particularly strong in weak fields, i.e., the low field magnetic reversibility is not due to vortex motion but rather because grains are too small to capture even a single vortex line. Indeed, if the size of the grain in the direction perpendicular to the field is smaller or of the order of the "size" of the magnetic flux quantum, $D_0 = \sqrt{\Phi_0/H}$ ($\Phi_0 = 2.05\cdot 10^{-7}$ G$\cdot$cm$^2$ is the magnetic flux quantum), no vortices can be created. As an estimate we can use the result of Ref. \onlinecite{doria} where it was shown that the first vortex is created when the transverse size $D\sim 2D_0$. A similar result was also obtained for very small superconducting spheres.\cite{bael} Although Refs. \onlinecite{doria} and \onlinecite{bael} are related to a somewhat different geometries, the difference with our case is not expected to be too big.

Considering data presented in Fig. 4, we may conclude that in $H = 1$ Oe there are practically no grains that can capture a vortex, while the magnetic field of 30 Oe is already strong enough to create the mixed state in a considerable number of grains. The values of $D_0$ are equal to 4.5 $\mu$m, 2.6 $\mu$m, and 0.8 $\mu$m for magnetic fields of 1 Oe, 3 Oe, and 30 Oe, respectively. This gives an estimate $r_{eff} \approx 3-5$ $\mu$m for the biggest grains and $r_{eff} \gtrsim 1$ $\mu$m for a large number of grains ($r_{eff}$ is an effective grain radius in the direction perpendicular to the magnetic field). 

\begin{figure}[t]
 \begin{center}
  \epsfxsize=1.0\columnwidth \epsfbox {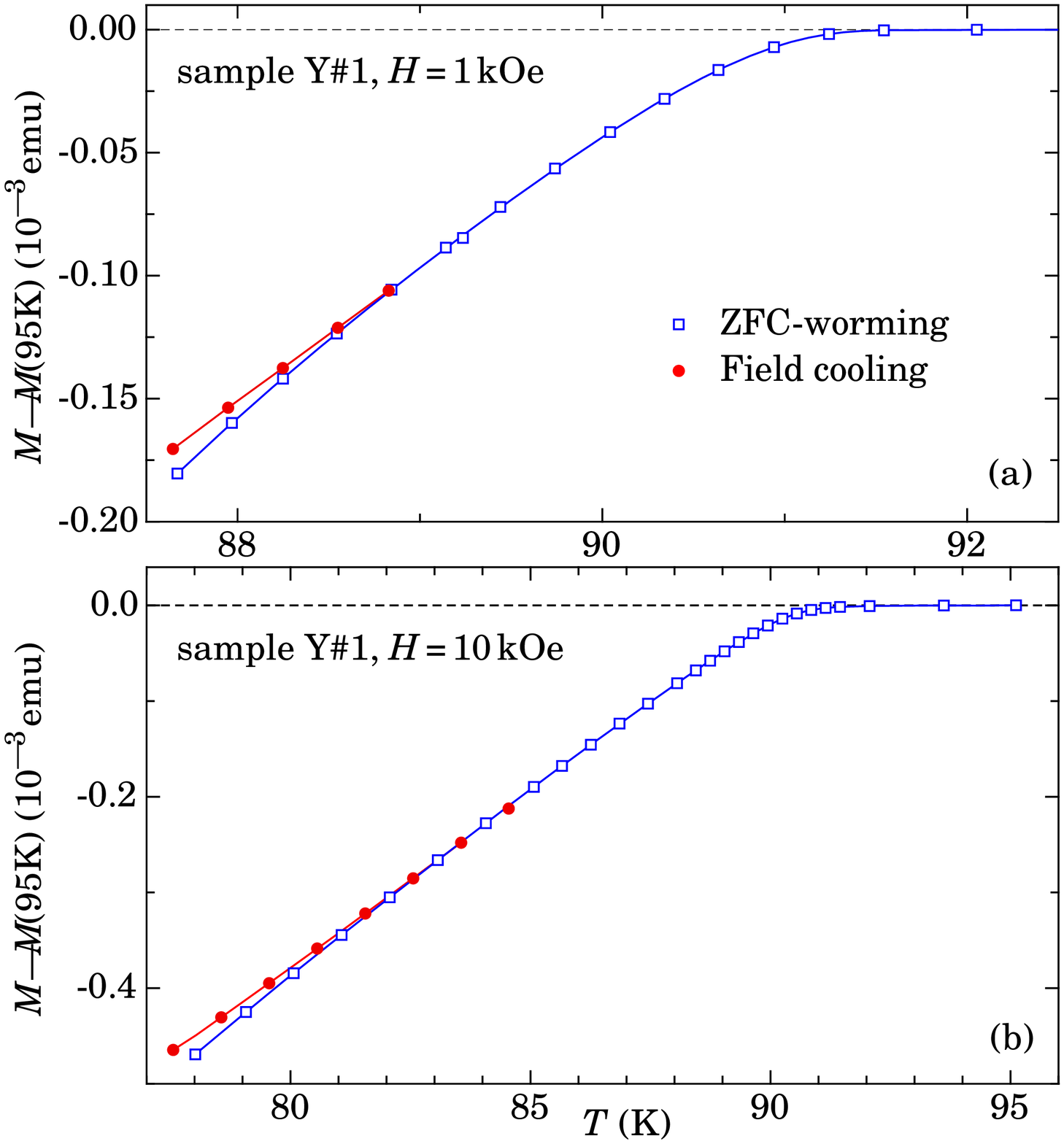}
  \caption{(Color online) ZFC and FC magnetization curves measured in (a) $H = 1$ kOe and (b) $H = 10$ kOe. The normal state magnetization at $T = 95$ K was subtracted. The solid lines are guides to the eye.}
 \end{center}
\end{figure}
In magnetic fields $H \gtrsim 1$ kOe the mixed state is created in practically all grains and one can see an extended range of reversible magnetization, as it is illustrated in Fig. 5, in which the difference $M(T) - M(95$K$)$ is potted. In the following analysis we shall use these equilibrium magnetization data in order to evaluate a temperature dependence of the normalized upper critical field by employing the scaling procedure introduced in Ref. \onlinecite{land1}.

\section{Analysis of experimental data}

\subsection{Superconducting grains in weak magnetic fields}

The magnetic moment of a small superconducting sample in the Meissner state depends on the ratio $\lambda/r$ where $2r $ is the size of the sample in the direction perpendicular to an applied magnetic field. This is why low field magnetization measurements may serve as a valuable tool for evaluation of $\lambda$ and its temperature dependence. Because the magnetic moment of a single sample with $r \sim \lambda$ is too small to be measured, powder or grain aligned samples were used.\cite{monod,panago1,panago2,panago3,porch,khas1,khas2,khas3,khas5,zuev} Here, we apply a similar approach to analyze magnetization data collected on ceramic samples. 

For a superconducting sphere the normalized magnetization is\cite{shoen}
\begin{equation}
\frac{M(\lambda/r)}{M_0} = 1 - 3\frac{\lambda}{r}\coth\frac{r}{\lambda} + 3\frac{\lambda^2}{r^2},
\end{equation}
where $M_0 = M(\lambda=0)$. Because the ratio $\lambda /r$ enters Eq. (2) in a rather complex way, the grain size distribution may be important. Usually, this distribution is measured independently and Eq. (2) is integrated over all grain sizes.\cite{porch} In our case, this could not be done reliably. Instead of this, we have analyzed, how the grain size distribution may influence magnetization results.

\begin{figure}[h]
 \begin{center}
  \epsfxsize=1.0\columnwidth \epsfbox {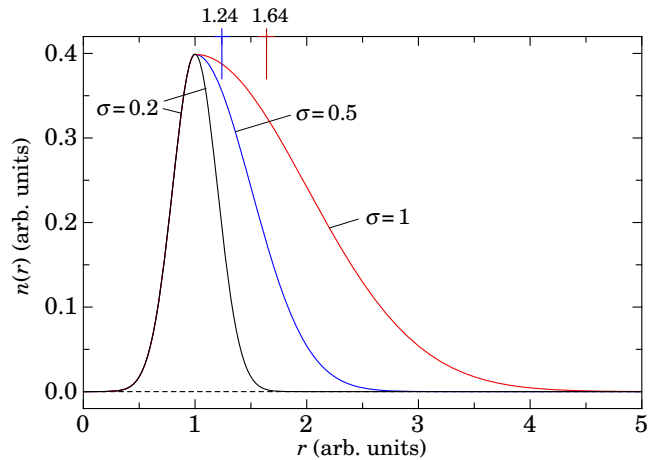}
  \caption{(Color online) Grain size distributions according to Eq. (3). The vertical lines indicate average values of radii ($r_0$) for two asymmetric distributions.}
 \end{center}
\end{figure}
Three different grain size distributions, as shown in Fig. 6, were used. In order to simplify the calculations, the following $n(r)$ were assumed:
\begin{equation}
n(r) = \frac{\pi}{2}\exp\bigg\{- \frac{(r - 1)^2}{2\sigma}\bigg\}
\end{equation}
with $\sigma = \sigma_1$ for $r \le 1$ and $\sigma = \sigma_2$ for $r >1$ (Fig. 6).

\begin{figure}[h]
 \begin{center}
  \epsfxsize=1.0\columnwidth \epsfbox {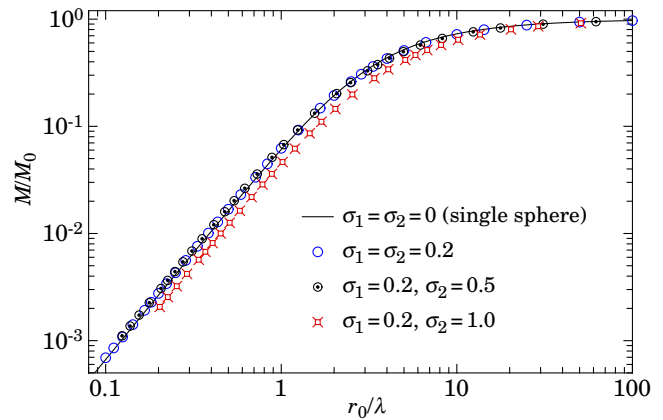}
  \caption{(Color online) The normalized magnetic response $M/M_0$ as a function of $r_0/\lambda$ calculated for model distributions described by Eq. (3) and presented in Fig. 6. The solid line shows the result for a single sphere.}
 \end{center}
\end{figure}
The results of calculations are presented in Fig. 7 on double logarithmic scales. Only in the case of the widest grain size distribution ($\sigma_2 =1$), the $M(r_0/\lambda)$ curve deviates noticeably from that for a single sphere. Furthermore, the difference between the curves can practically be eliminated by a parallel shift of the curve, which corresponds to some renormalization of $r_0$ and $M_0$ (see Fig. 7). In other words, one can replace integration of Eq. (2) by introduction of an effective $r_0$, which is close to the average grain radius. This replacement may lead to errors in the absolute value of $\lambda$, but it is not affecting its temperature dependence. 

\subsection{Magnetization in low magnetic fields}

Eq. (2) gives $M$ as a function of $\lambda/r_0$, while experiments provide the $M(T)$ curve. In the present analysis, we assume that at temperatures, close to $T_c$, the temperature dependence of $\lambda$ follows predictions of the Ginzburg-Landau theory:
\begin{equation}
\lambda(T) = \lambda^{(0)}_{GL}/\sqrt{1 - \tau}
\end{equation}
with $\tau = T/T_c$. This expression is rather general and it should be valid for any superconductor at temperatures sufficiently close to $T_c$. We underline that $\lambda^{(0)}_{GL}$ is a parameter describing the temperature dependence of $\lambda$ at $(1- \tau) \ll 1$ and does not represent the value of the magnetic field penetration depth at $T=0$. 

\begin{figure}[h]
 \begin{center}
  \epsfxsize=1.0\columnwidth \epsfbox {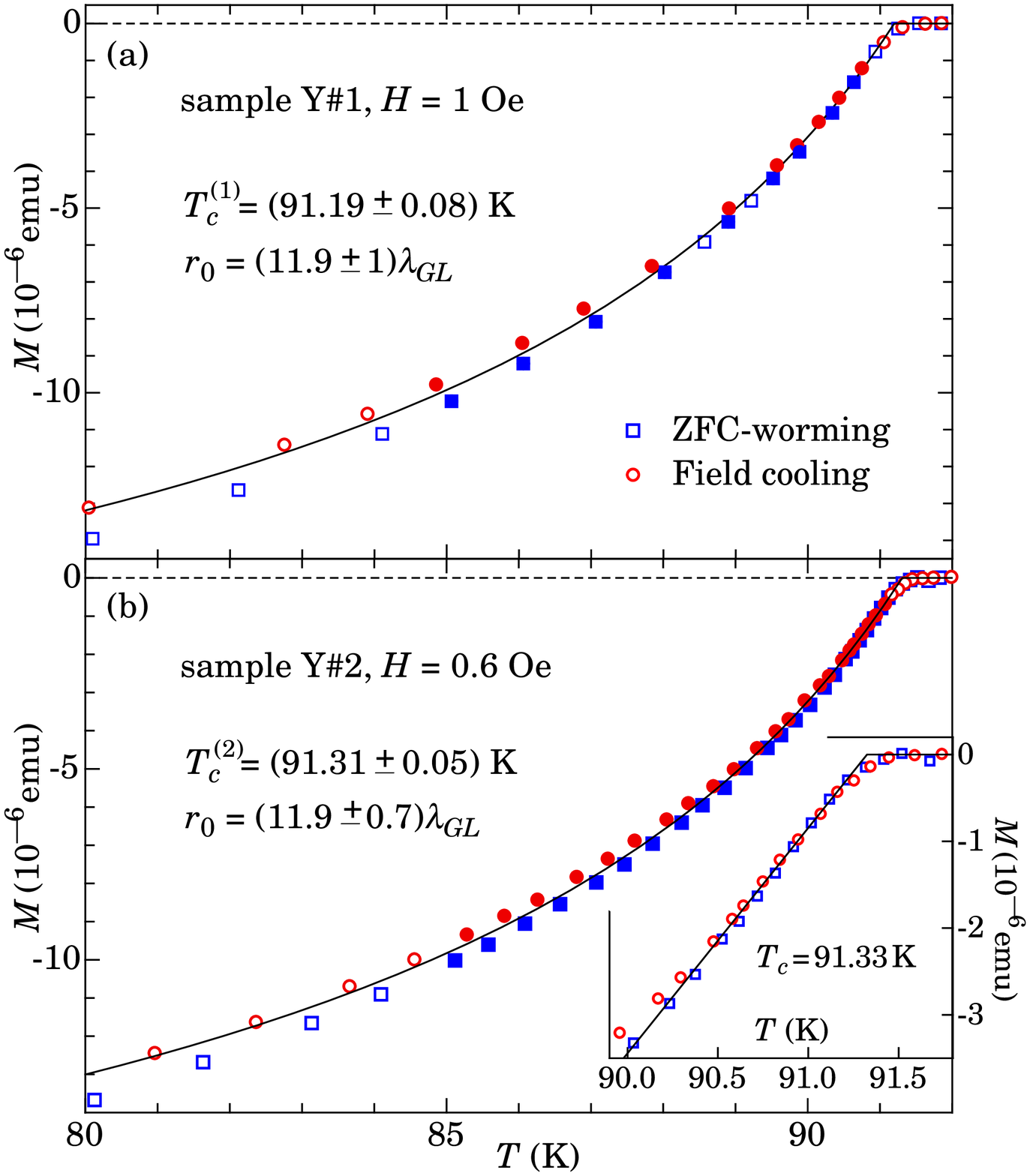}
  \caption{(Color online) High temperature parts of the $M(T)$ curves measured in ultra low fields. (a) Sample Y\#1. (b) Sample Y\#2. The solid lines are fits of Eq. (5) to experimental data for $T \ge 85$ K. Only data points marked by closed symbols were used for fitting. The values of $T_c$ and $r_0$ resulting from our analysis are indicated in the figure. The inset shows $M(T)$ data at temperatures very close to $T_c$. The straight line is the best linear fit to data points in the temperature range $90.5$ K $< T < 91.2$ K. This fit results in $T_c = (91.33 \pm 0.03)$ K.}
 \end{center}
\end{figure}
Substituting Eq. (4) in Eq. (2) and introducing $\Lambda = \lambda^{(0)}_{GL}/r_0$, we obtain
\begin{equation}
\frac{M(T)}{M_0} =1 - \frac{3\Lambda}{\sqrt{1 - \tau}} \left({\coth\frac{\sqrt{1 - \tau}}{\Lambda}} - \frac{\Lambda}{\sqrt{1 - \tau}} \right).
\end{equation}
Eq. (5) may straightforwardly be employed for the analysis of experimental $M(T)$ data. There are three adjustable parameters in Eq. (5), which are not known {\it a priori}. $M_0$ changes the scale of the $M(T)$ curve along the vertical axis, the value of $T_c$ is the temperature, at which the diamagnetic moment of the sample vanishes, and $\Lambda = \lambda^{(0)}_{GL}/r_0$ describes the curvature of the $M(T)$ curve. Because these parameters are related to rather different characteristics of the curve, all of them can reliably be evaluated. We also note some rounding of the $M(T)$ curve at temperatures very close to $T_c$ (see inset of Fig. 8(b)). The smearing of the transition is weak and may be seen only in a rather narrow temperature range near $T_c$. The corresponding data points were not used in the analysis.  

As it is demonstrated in Fig. 8, Eq. (5) provides a rather good approximation to experimental data at temperatures $T \ge 85$ K. This allows for a precise evaluation of the superconducting critical temperature for both samples with $T^{(1)}_c = (91.19 \pm 0.08)$ K (Y\#1) and $T^{(2)}_c = (91.31 \pm 0.05)$ K (Y\#2). Amazingly, the parameter $\Lambda$ turned out to be identical for both samples (see Fig. 8). Because the samples were prepared under the same conditions, identical grain size distributions are expected. At the same time, very close values of $r_0$ for both samples obtained as a result of the analysis of magnetization data may be considered as a confirmation of the validity of this approach. 

According to Eq. (2), the magnetic moment is inversely proportional to $\lambda^2$  if $\lambda(T) > r$. Taking into account that the temperature dependence of $\lambda$ is expressed by  Eq. (4), we get $M \sim (1 - T/T_c)$. Because $\lambda$ diverges at $T = T_c$, the condition $\lambda(T) > r$ is always fulfilled in the vicinity of $T_c$. This simple conclusion is in agreement with experimental results presented in the inset of Fig. 8(b). As may be seen, at temperatures 90.5 K $< T < 91.2$ K, $M(T)$ data points may very well be approximated by a straight line. Linear extrapolation to $M = 0$ gives $T_c = (91.33 \pm 0.03)$ K, which is in excellent agreement with $T_c = (91.31 \pm 0.05)$ K obtained by fitting of experimental data with Eq. (5) in a much wider temperature range.

\subsection{Magnetization in high magnetic fields}

\begin{figure}[t]
 \begin{center}
  \epsfxsize=1.0\columnwidth \epsfbox {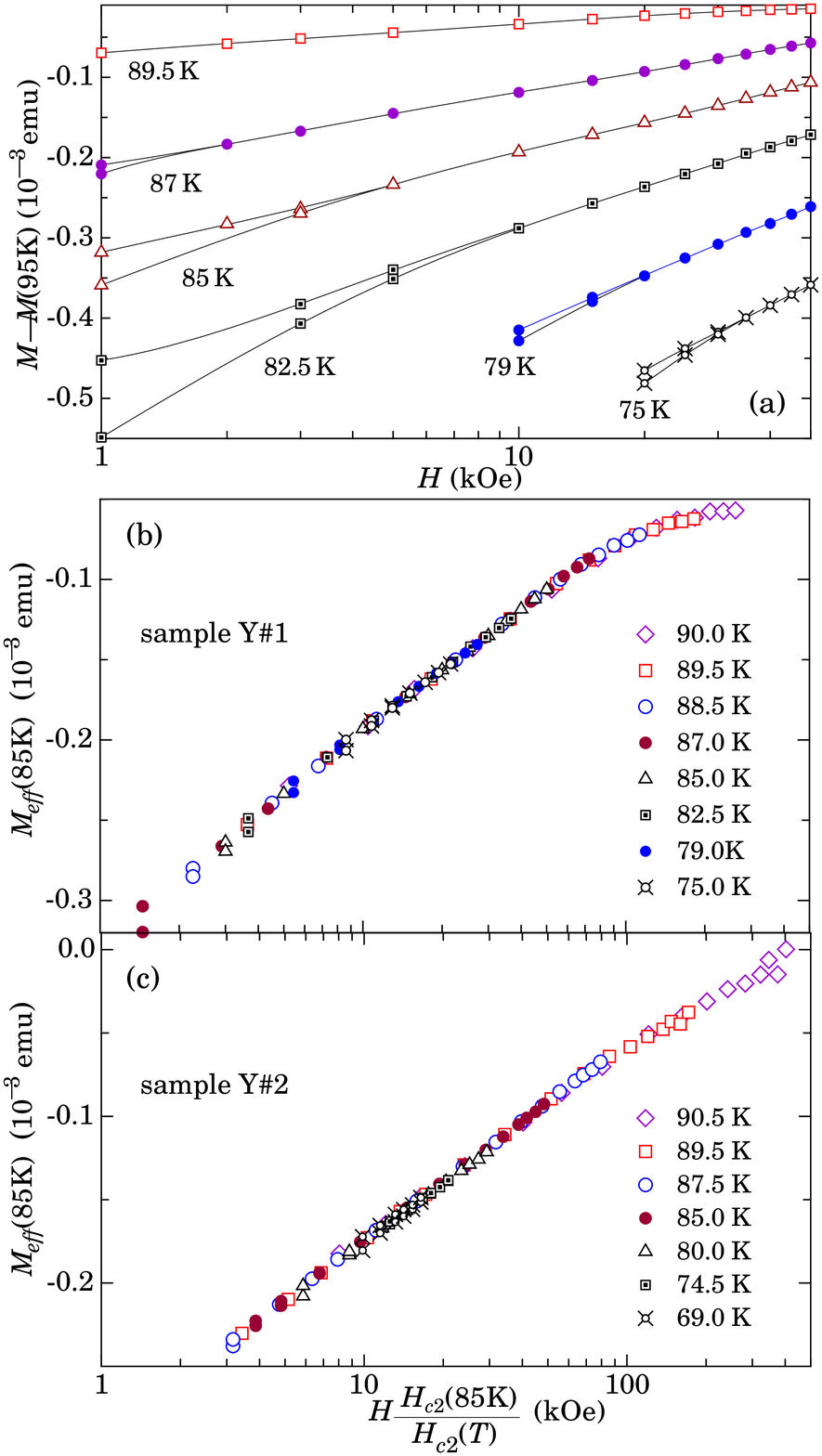}
  \caption{(Color online) (a) Examples of the $M(H)$ curves for the sample Y\#1. The solid lines are guides to the eye. (b) and (c)  $M_{eff}(85$K$)$ as a function of the normalized field for samples Y\#1 and Y\#2, respectively. Different symbols correspond to $M_{eff}$ calculated from $M(H)$ data collected at different temperatures.}
 \end{center}
\end{figure}
Here we use reversible magnetization data collected in magnetic fields 1 kOe $\le H \le 50$ kOe in order to evaluate the temperature dependence of the normalized upper critical field $H_{c2}$. A  scaling procedure developed in Ref. \onlinecite{land1} was used for the analysis of experimental data.

The scaling procedure is based on a single assumption that the Ginzburg-Landau parameter $\kappa$ is temperature independent. In this case, magnetizations measured at different temperatures but in the same normalized fields $H/H_{c2}(T)$ are proportional to $H_{c2}(T)$. According to Ref. \onlinecite{land1}, the magnetizations of a sample at two different temperatures $T$ and $T_0$ are related by
\begin{equation}
M(H,T_0)=M(h_{c2}H,T)/h_{c2}+c_0(T)H,
\end{equation}
where $h_{c2} = H_{c2}(T)/H_{c2}(T_{0})$ is the normalized uper critical field and $c_{0}(T)= \chi_{n}(T_{0}) - \chi_{n}(T)$ ($\chi_{n}$ is the normal-state magnetic susceptibility of a sample). It is important to underline that in experiments $\chi_n$ includes also a contribution arising from the sample-holder, which can be substantial in the case of small samples. While the first term on the right side of Eq. (6) describes the properties of the mixed state of ideal type-II superconductors, the second one is introduced in order to account for all other temperature dependent contributions to magnetization. By a suitable choice of $h_{c2}$ and $c_{0}(T)$ individual $M(H)$ curves measured at different temperatures may be merged into a single master curve $M_{eff}(H,T_0)$. In this way the temperature dependence of the normalized upper critical field can be obtained.\cite{land1} Any temperature inside the investigated temperature range may be chosen as $T_0$. $M_{eff}(H)$ represents the equilibrium magnetization curve for $T = T_0$. It should be noted that in addition to the diamagnetic response of the mixed state $M_{eff}$ also includes a $\chi_n(T_0)H$ term.

We measured $M(T)$ for different magnetic fields, as shown in Fig. 5. $M(T)$ data can easily be transformed into $M(H)$ curves. Several such curves are shown in Fig. 9(a). In the analysis below we use $[M - M(95$K$)]$ instead of $M$. Because only the difference $[\chi_n(T_0) - \chi_n(T)]$ enters $c_0(T)$, such subtraction does not change the scaling procedure.  On the other hand, subtraction of $M(95$K$)$ reduces non-superconducting contribution to $M_{eff}$.

Figs. 9(b) and 9(c) show the scaled magnetization curves for both investigated samples. As may be seen, agreement between values of $M_{eff}$ calculated from data collected at different temperatures is practically perfect in both cases. The curves shown in Figs. 9(b) and 9(c) represent the equilibrium magnetization curves for $T = 85$ K. While direct measurements may provide such curves in magnetic fields 3 kOe $\le H\le 50$ kOe (lower field data are unavailable because of irreversibility), the scaling procedure allows to establish the magnetization curves in a 10 times wider magnetic field range. 

\begin{figure}[h]
 \begin{center}
  \epsfxsize=1.0\columnwidth \epsfbox {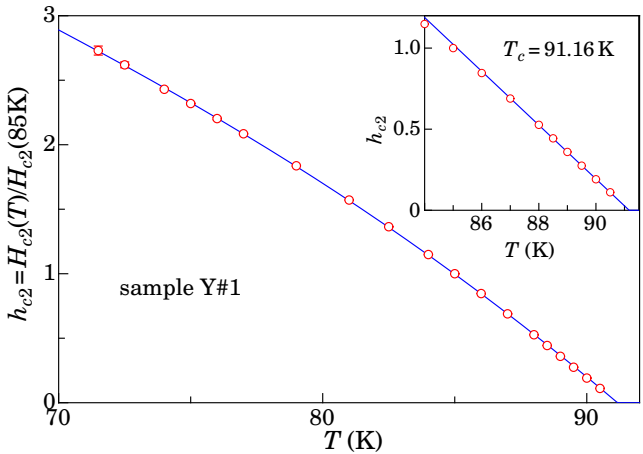}
  \caption{(Color online) The normalized upper critical field versus temperature. The solid line is a guide to the eye. The inset shows the region near $T_c$ on expanded scales. The solid line in the inset is the best linear fit to data points at $T \ge 87$ K. The resulting value of $T_c$ is indicated in the figure.}
 \end{center}
\end{figure}
Fig. 10 shows the resulting $h_{c2}(T)$ dependence for the sample Y\#1. As may be seen in the inset of Fig. 10, $h_{c2}$ is a linear function of temperature at $T \ge 87$ K. This linearity allows for a rather accurate evaluation of $T_c$ by extrapolation of $h_{c2}(T)$ to $h_{c2} = 0$. As a result we obtain $T^{(1)}_c =(91.16 \pm 0.05)$ K. This value is practically the same as $T^{(1)}_c$ evaluated from low field $M(T)$ measurements (see Fig. 8(a)). A similar procedure for the sample Y\#2 results in $T^{(2)}_c =(91.26 \pm 0.05)$ K. This value is again in very good agreement with the result of low field measurements (see Fig. 8(b)). 

\begin{figure}[h]
 \begin{center}
  \epsfxsize=1.0\columnwidth \epsfbox {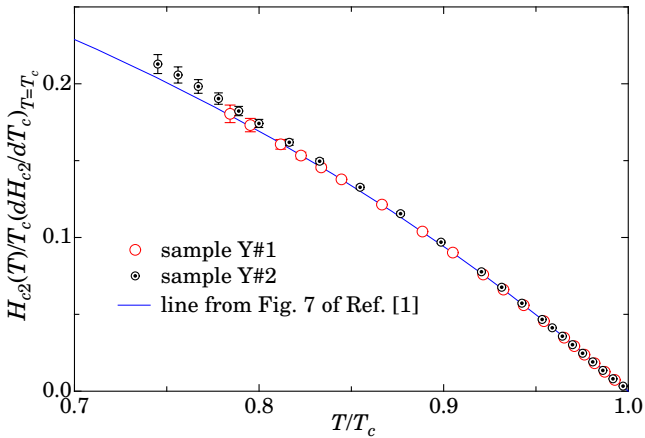}
  \caption{(Color online) The upper critical field normalized by $T_c(dH_{c2}/dT)_{T=T_c}$ as a function of $T/T_c$ for two samples. The dashed line  was obtained in Ref. \onlinecite{land1} by the analysis of magnetization data for series of Bi-based cuprates.}
 \end{center}
\end{figure}
Because the two sample have slightly different critical temperatures, we plot in Fig. 11 the normalized upper critical field as a function of $T/T_c$. For this graph, $H_{c2}$ was normalized by $T_c(dH_{c2}/dT)_{T=T_c}$. As may be seen, the curves for two samples perfectly match each other. 

It was established earlier that considering the $H_{c2}(T)$ curves, numerous high-$T_c$ superconductors may be divided into two groups with practically identical normalized $H_{c2}(T)$ curves for each of the group.\cite{land1,land2,land3,land4} The dashed line in Fig. 11, which is taken from Ref. \onlinecite{land1}, represents such a curve for the larger group. As may be seen in Fig. 11, present results for ceramic samples are fully consistent with this curve. We argue that this is a rather strong indication that the normalized $H_{c2}(T)$ curves are practically independent of the orientation of the magnetic field.

\section{Discussion}

\subsection{Magnetization in high magnetic fields}

As it was demonstrated in Section II, scaling of the $M(H)$ curves allows to establish temperature dependencies of the normalized upper critical field (Figs. 10 and 11) and to evaluate the superconducting critical temperature for each sample as a point, at which $H_{c2}$ vanishes (see inset of Fig. 10). The value of $T_c$ can also be evaluated from low field $M(T)$ measurements, as it is demonstrated in Fig. 8. The comparison of these two analyses are presented in Table I.

\begin{table}[h]
\caption{Summary of evaluation of $T_c$ }
\begin{tabular}{lcccccc}
\multicolumn{1}{c}{ evaluation method } &
\multicolumn{1}{c}{sample Y\#1} &
\multicolumn{1}{c}{sample Y\#2} & \\

$T_c$ from low field $M(T)$ data  & $91.19 \pm 0.08$ & $91.31 \pm 0.05$ &    \\
$T_c$ from $h_{c2}(T)$ curves & $91.16 \pm 0.05$ & $91.26 \pm 0.05$  & \\ 

\end{tabular}
\end{table}
We emphasize that $T_c$ was evaluated from two completely different sets of experimental data, which correspond to different physical processes. Indeed, the low-field evaluation of $T_c$ is obtained from the temperature dependence of the magnetic field penetration depth, while the high-field estimate is based on the temperature dependence of $H_{c2}$. We argue that close agreement between the two is convincing evidence that both approaches correctly interpret experimental results.

Error margins indicated in Table I are statistical errors evaluated by a fit-procedure and do not include possible systematic errors of theoretical approaches. As may be seen in Table I, $T_c$ evaluated from the $h_{c2}(T)$ curves is slightly below the low-field estimate. The difference, however, is two small to be speculated on.

The scaling procedure, which was used for the analysis of magnetization data is based on the assumption that the Ginzburg-Landau parameter $\kappa$ is temperature independent, which is not necessary correct. This is why recent direct measurements $H_{c2}(T)$ in magnetic fields up to 600 T are of primary importance.\cite{sekitani} Comparison of our normalized $H_{c2}(T)$ curve, presented in Fig. 11, with the results of Ref. \onlinecite{sekitani} shows a very good agreement and justifies our choice of temperature independent $\kappa$.

\subsection{The grain size and the $r_0/\lambda$ ratio.}

As was already discussed (Sections IV and IV), magnetization measurement allows for evaluation of $\lambda/r_0$ where $r_0$ is the effective grain radius in the direction perpendicular to the magnetic field. Because in our case the grains have irregular shapes and they are not oriented, reliable estimations of the absolute value of $\lambda$ are not feasible. We can, however, move in the opposite way and use literature values of $\lambda$ in order to evaluate $r_0$. This is especially interesting because, as we discuss below, magnetization measurements provide three independent ways for evaluation of $r_0$ and comparison of the resulting values may serve as a consistency check of the theoretical approach. In order to distinguish the different evaluations of $r_0$, we shall use upper indexes.

The first estimate of the grain size, discussed in Section III, is independent of the magnetic field penetration depth and stems from the fact that in order that the mixed state in separated grains can be created, their size in the direction perpendicular to the magnetic field must be about twice larger than the size of the magnetic flux quantum $D_0 = \sqrt{\Phi_0/H}$.\cite{doria,bael} If grains are too small, no vortices can be created and the sample magnetization $M(T)$ has to be reversible. In the opposite case, $M(T)$ is irreversible. Taking into account that the $M(T)$ curves measured in fields $H \le 3$ Oe are practically reversible, while those in fields $H\ge 30$ Oe are clearly irreversible (see Fig. 4), we obtain $r^{(1)}_0 > 1$ $\mu$m (see Section III). 

The second estimate was obtained from the fit of the $M(T)$ curve at $T > 85$ K (see Fig. 8). This gives $r_0 = 11.9 \lambda^{(0)}_{GL}$. The value of $ \lambda^{(0)}_{GL}$ does not have real physical meaning and it cannot be measured, however, using Eq. (4), we can replace $ \lambda^{(0)}_{GL}$ by a measurable value of $\lambda$ related to some particular temperature. As a reference value we have chosen $\lambda(0.95T_c)$. The effective grain radius may then be written as $r_0 = 2.7 \lambda(0.95T_c)$.

Considering $\lambda/r_0$ in such anisotropic materials as Y-123, the anisotropy of $\lambda$ has to be taken into account. In Y-123, the value of $\lambda$ for currents flowing in the $c$-direction $\lambda_c$ is about 7 times larger than  $\lambda_{ab}$.\cite{zheng} At higher temperatures, at which $\lambda_c > r_0$, the situation is simplified by a $1/\lambda^2$ dependence of $M$ (see Eq. (2)). In this case, the main contribution to $M$ arises from  grains with $ab$-planes approximately perpendicular to $H$. In this temperature range, averaging leads to $\lambda \approx 1.8 \lambda_{ab}$. Taking $\lambda_{ab}(0.95T_c) = 0.33$ $\mu$m,\cite{kamal1,kamal2} we obtain $r^{(2)}_0 \approx 1.6$ $\mu$m in good agreement with the previous estimate. We underline that this is a rather approximate result. Averaging was done assuming that the grain size in the direction perpendicular to the magnetic field does not depend on the crystallographic orientation of the grain, which is not exactly true.

In fact, there is one more way to evaluate the grain size. As is demonstrated in Fig. 3, there is a magnetic field range, in which the sample magnetization is reversible at low temperatures. This reversibility has the same origin as reversibility in very low fields and high temperatures (Fig. 8). In this magnetic field range, there are no vortices inside grains and the difference between the ideal Meissner state ($\chi = -1/4\pi$) and the sample magnetization is due to a non-zero value of $\lambda$.  In other words, magnetization measurements at low temperatures and in magnetic fields of the order of 100 Oe may also be used to obtain information about the $r_0/\lambda$ ratio. 

At low temperatures, both $\lambda_c$ and $\lambda_{ab}$ are sufficiently small and, according to Eq. (2), averaging is dependent on the actual value $r_0$. In the following we take $r_0 = 1.6$ $\mu$m as it was estimated above.  Taking $\lambda_c = 7\lambda_{ab}$ and using the commonly accepted value of $\lambda_{ab}(0$K$) = 0.15$ $\mu$m,\cite{lambda} we obtain $\lambda \approx 2.6 \lambda_{ab}$.

This estimation of the grain size is the only result of this work, which relies on the absolute value of $M$. For the evaluation of $r_0/\lambda$ we need to know a demagnetizing factor $N_{eff}$ of our sample consisting of a large number of non-intracting grains. $N_{eff}$ is a result of some averaging of demagnetizing factors of individual grains and it cannot be evaluated experimentally. For the folowing estimate we take $N_{eff} = 1/3$. From the slope $dM/dH$ calculated for data presented in Fig. 3, we obtain $M/M_0 = 0.39$. Substituting this value in Eq. (3), we get $r_0= 3.6\lambda_{eff}$. Using the value of $\lambda_{eff}$ evaluated in the previous paragraph, we obtain $r^{(3)}_0 \approx 1.4$ $\mu$m. Although this value of $r_0$ was obtained from completely different experimental data, it is in amazingly good agreement with $r^{(2)}_0$. 

\begin{table}[h]
\caption{ Summary of $r_0$ results}
\begin{tabular}{lcccccc}
\multicolumn{1}{c}{evaluation  method } &
\multicolumn{1}{c}{ $r_0$  } & \\

onset of irreversibility & $ > 1$ $\mu$m \\
$M(T)$ at $T > 85$ K & 1.6 $\mu$m \\
$dM/dH$ at $T = 4.5$ K & 1.4 $\mu$m \\

\end{tabular}
\end{table}
The resulting values of $r_0$ are summarized in Table II. While all three values are in reasonable agreement with each other, they have rather different uncertainties. A transition from the reversible to irreversible behavior with increasing magnetic field gives only a range of $r_0$ values. In the evaluation of the grain size from low temperature $M(H)$ measurements ($r^{(3)}_0$) enters an unknown value of $N_{eff}$ introducing an additional uncertainty. This is why we consider $r^{(2)}_0 = 1.6$ $\mu$m as the most reliable. 

\begin{figure}[h]
 \begin{center}
  \epsfxsize=1.0\columnwidth \epsfbox {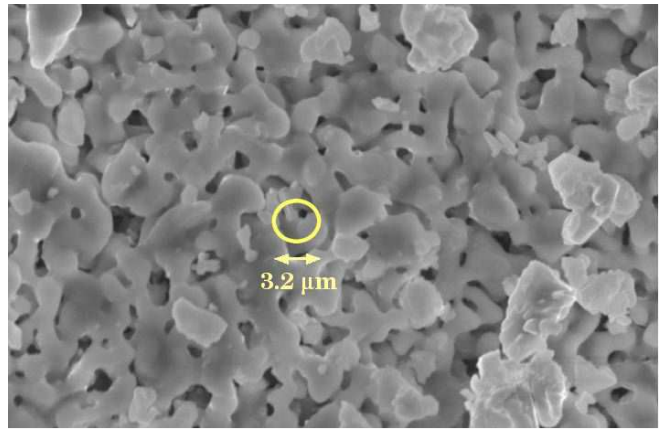}
  \caption{(Color online) Micrograph of one of the samples. The circle radius corresponds to our estimation of the grain size ($2r_0 = 3.2$ $\mu$m).}
 \end{center}
\end{figure}
Fig. 12 shows a micrograph of one of spheres, which was intentionally broken. The circle diameter corresponds to $r^{(2)}_0$. Taking into account an approximate character of the consideration, this agreement must be considered as more than satisfactory. We argue that this agreement represents convincing evidence that our approach to the analysis of experimental data is generally correct.

\subsection{The temperature dependence of $\lambda$}

While the absolute values of $\lambda$ can hardly be evaluated from data collected on ceramic samples, the fact that at temperatures close to $T_c$ K $\lambda(T)$ follows predictions of the Ginzburg-Landau theory is established quite reliably. This is illustrated in Fig. 8. As may be seen, $M(T)$ data points follow the theoretical curve at $T > 85$ K. Furthermore, at temperatures very close to $T_c$, the $M(T)$ dependence is perfectly linear (see inset of Fig. 8(b)) as it is expected from the theory.  

Considering this result it must be clearly understood that our consideration is only applicable in the vicinity of $T_c$. There is no way to make any definite conclusion about $\lambda(T)$ at lower temperatures if ceramic samples are used. This can only be done in experiments on grain-aligned samples.

As was discussed in the previous subsection, because at these temperatures $\lambda_c \gg \lambda_{ab} \sim r_0$ and $M \sim 1/\lambda^2$, the contribution to the sample magnetization arising from grains with $ab$-planes oriented along the field is negligibly small and can be disregarded. This means that our results for $\lambda(T)$ are related to $\lambda_{ab}$. 

The $\lambda_{ab}(T)$ dependence obtained in this work is in agreement with some studies\cite{monod,panago1,panago2,plus1a,plus1,plus2,plus3,plus4} and in disagreement with data from others.\cite{kamal1,kamal2,minus} We point out that the behavior described by Eq. (4) was observed for grain aligned samples,\cite{panago1,panago2} films,\cite{plus1a,plus1,plus2} and single crystals.\cite{plus3,plus4} At the same time, the results of Refs. \onlinecite{kamal1,kamal2, minus} provide $\lambda_{ab} \sim 1/(1-T/T_c)^{1/3}$ instead of Eq. (4). As far as we are aware this controversy is still unresolved. We may note that in Ref. \onlinecite{kamal2} the $1/(1-T/T_c)^{1/3}$ dependence of $\lambda_{ab}$ was extended up to $T \sim 0.9996T_c$. This cannot be really justified. Indeed, the width of the superconducting transition makes evaluation of $T_c$ with such an accuracy impossible. We argue that even a 0.1 K error in $T_c$ can substantially distort the $\lambda(T)$ dependence at $T > 0.99T_c$. 

\section{Conclusion}

It was demonstrated that low density ceramic samples of YBa$_2$Cu$_3$O$_{7-x}$ in certain conditions may be considered as systems of non-intracting grains. It was also shown that low-field magnetization measurements on such samples provide three independent ways for an evaluation of the grain size (see Table II). All three values are in good agreement with each other and, what is more important, the evaluation of $r_0$ is in good agreement with the real sample structure (Fig. 12).

At temperatures close to $T_c$, low-field magnetization data may very well be described assuming that the temperature dependence of $\lambda$ follows Eq. (4), which is the result of the Ginzburg-Landau theory (Fig. 8). Although ceramic samples with non-oriented grains were used, it was possible to demonstrate that the above mentioned result is related to $\lambda_{ab}$ (see Section V).

The analysis of high field magnetization data allowed to establish the temperature dependence of the normalized upper critical field, which is in agreement with previous results obtained on numerous high-$T_c$ superconducting compounds. In this work, the upper limit of the investigated temperature range was extended up to $T \approx 0.995T_c$ and it was demonstrated that the $H_{c2}(T)$ dependence is a linear function of $T$ at $T \gtrsim 0.95T_c$. This result is in agreement with the Ginzburg-Landau theory. We note that both $H_{c2}$ and $\lambda_{ab}$ follow the Ginzburg-Landau theory in approximately the same range of temperatures.

Magnetization measurements presented in this work allowed for evaluation of $T_c$ from two completely different sets of experimental data. It turned out that the both results are in perfect agrement (see Table I).

\section{ACKNOWLEDGMENTS}

This work was in part supported by he NCCR MaNEP-II of the Swiss National Science Foundation (Project 4).

\end{document}